# Symmetry-break, mixing, instability, and low frequency variability in a minimal Lorenz-like system

Valerio Lucarini

Department of Physics, University of Bologna, Viale Berti-Pichat 6/2 40127 Bologna, Italy

Istituto Nazionale di Fisica Nucleare, Sezione di Bologna, Via Irnerio 46, 40127 Bologna, Italy

Klaus Fraedrich

Institute of Meteorology, University of Hamburg, Bundestrasse 55, 20146 Hamburg, Germany

CLISAP, University of Hamburg, Grindelberg 5, 20144, Hamburg, Germany

**Abstract**

Starting from the classical Saltzman 2D convection equations, we derive via a severe spectral truncation a minimal 10 ODE system which includes the thermal effect of viscous dissipation. Neglecting this process leads to a dynamical system which includes a decoupled (generalized) Lorenz system. The consideration of this process breaks an important symmetry, couples the dynamics of fast and slow variables, ensuing modifications of the structural properties of the attractor and of the spectral features. When the relevant nondimensional number (Eckert number $Ec$) is different from zero, the system is ergodic and hyperbolic, the slow variables feature long term memory with $1/f^{3/2}$ power spectra, and the fast variables feature amplitude modulation on time scale of $1/Ec$. Increasing the strength of the thermal-viscous feedback has a stabilizing effect, as both the metric entropy and the Kaplan-Yorke attractor dimension decrease monotonically with $Ec$. The analyzed system features very rich dynamics: it overcomes some of the limitations of the Lorenz system and might have prototypical value in relevant processes in complex systems dynamics, such as the interaction between slow and fast variables. the presence of long term memory and the associated extreme value statistics. Analysis shows how, neglecting the coupling of slow and fast variables only on the basis of scale analysis can be catastrophic. In fact, this leads to spurious invariances that affect essential dynamical properties (ergodicity, hyperbolicity) and that cause the model losing ability in describing intrinsically multiscale processes.

**Introduction**

The Lorenz system (Lorenz, 1963) has a central role in modern science as it has provided the first example of low-dimensional chaos (Ruelle, 1989), and has literally paved the way for new scientific paradigms. The Lorenz system can be derived with a minimal truncation of the Fourier-modes projection of the 2D Boussinesq convection equations introduced by Saltzman (1962), where a specific selection of the spatial symmetry of the fields is considered. Extensions of the Lorenz system taking into account higher-order spectral truncations have been presented, see e.g. Curry (1978), Wenyi and Peicai (1984), Roy and Musielak (2007).

The mathematical properties of this system have been the subject of an intense analysis, which has addressed the bifurcations as well as the characteristics of the strange attractors realized within certain parametric ranges. For several classic results, see Sparrow (1984). Recently, at more theoretical level, the investigation of Lorenz-like systems has stimulated the introduction of the family of singular hyperbolic systems as extension of the family of hyperbolic systems (Bonatti et al., 2005). Moreover, moving from the theory developed for non-equilibrium systems (Lucarini, 2008a) starting from the Ruelle response theory (Ruelle, 1998), in a recent paper Lucarini (2008b) has performed a careful verification on the Lorenz system of the Kramers-Kronig dispersion relations and sum rules.

In spite of its immense value, the Lorenz system does not provide an efficient representation of several crucial phenomena typically associated to complex, chaotic systems. When looking at finite time predictability properties, the Lorenz system features an unrealistic return-of-skill in the forecast, *i.e.* there are regions in the attractor within which all infinitesimal uncertainties decrease with time (Smith et al., 1999). Additionally, the Lorenz system does not feature an interplay between fast and slow variables, so that it cannot mimic the coupling between systems with different internal time scales. Palmer (1993) introduced artificially *ad hoc* components to the Lorenz system in order to derive a toy-model able to generate low-frequency variability. Moreover, as discussed in Nicolis (1999), the Lorenz system does not allow for a closed-form computation of the entropy production.

Some pathologies emerge also when looking at the relationship between the Lorenz system and higher order spectral truncation of the Saltzman equations. Recently, Chen and Price (2006) showed that the Lorenz system is a specific member of an class of equivalence of systems whose dynamics is invariant with respect to a symmetry transformation.

In this work, we wish to propose a minimal dynamical system, possibly with some paradigmatic value, able to overcome some of the limitation of the Lorenz system, and endowed with a much richer dynamics. Starting from the 2D convection equations, we derive with a truncation *à la Lorenz* a minimal 10 d.o.f. ODEs system which includes the description of the thermal effect of viscous dissipation. We discuss how, while neglecting this process lead us to a dynamical system which includes a decoupled (generalized) Lorenz system, its consideration breaks an important symmetry of the systems, couples the dynamics of fast and slow variable, with the ensuing modifications in the structural properties of the attractor and in the spectral features of the system.

**Rayleigh-Benard Convective System**

We consider a thermodynamic fluid having kinematic viscosity $\nu$, thermal conductivity $k$, thermal capacity at constant volume $C_v$, and linearized equation of state $\rho = \rho_0(1-\alpha T)$. The 2D $(x,z)$ Rayleigh–Benard top-heavy convective system confined within $z \in [0, H]$ can be completely described, when adopting the Boussinesq approximation, by the following system of PDEs:

$$\partial_t \nabla^2 \psi + J(\psi, \nabla^2 \psi) = g\alpha \partial_x \theta + \nu \nabla^4 \psi$$
$$\partial_t \theta + J(\psi, \theta) = \frac{\Delta T}{H} \partial_x \psi + k \nabla^2 \theta + \frac{\nu}{C_v} \partial_{ij}\psi \partial_{ij}\psi$$
(1a-1b)

where $g$ is the gravity acceleration, $\vec{v} = (u, w) = (-\partial_z \psi, \partial_x \psi)$, $T = T_0 - z\Delta T/H + \theta$, with $H$ uniform depth of the fluid and $\Delta T$ imposed temperature difference. The suitable boundary conditions in the case of free-slip system are $\theta(x, z=0) = \theta(x, z=H) = 0$ and $\psi(x, z=0) = \psi(x, z=H) = \nabla^2\psi(x, z=0) = \nabla^2\psi(x, z=H) = 0$ for all values of $x$. These PDEs can be non-dimensionalized by applying the linear transformations $x = H\hat{x}$, $z = H\hat{z}$, $t = H^2/k\,\hat{t}$, $\psi = k\hat{\psi}$, and $\theta = k\nu/(g\alpha H^3)\hat{\theta}$:

$$\partial_{\hat{t}} \hat{\nabla}^2 \hat{\psi} + J(\hat{\psi}, \hat{\nabla}^2 \hat{\psi}) = \sigma \partial_{\hat{x}} \hat{\theta} + \sigma \hat{\nabla}^4 \hat{\psi}$$
$$\partial_{\hat{t}} \hat{\theta} + J(\hat{\psi}, \hat{\theta}) = R \partial_{\hat{x}} \hat{\psi} + \hat{\nabla}^2 \hat{\theta} + \sigma Ec\, \partial_{\hat{ij}}\hat{\psi} \partial_{\hat{ij}}\hat{\psi}$$
(2a-2b)

where $\sigma = \nu/k$ is the Prandl number, $R = g\alpha H^3 \Delta T/(k\nu)$ is the Rayleigh number, and $Ec = k^2/(C_v \Delta T H^2)$ is the Eckert number, and the suitable boundary conditions are $\hat{\theta}(\hat{x}, \hat{z}=0) = \hat{\theta}(\hat{x}, \hat{z}=1) = 0$ and $\hat{\psi}(\hat{x}, \hat{z}=0) = \hat{\psi}(\hat{x}, \hat{z}=1) = \hat{\nabla}^2\hat{\psi}(\hat{x}, \hat{z}=0) = \hat{\nabla}^2\hat{\psi}(\hat{x}, \hat{z}=1) = 0$. The Eckert number in Eq. 2b quantifies the impact of viscous dissipation on the thermal balance of the system. As this number is usually rather small in actual fluids, the corresponding term in the previous set of PDEs is typically discarded. Following the strategy envisioned by Saltzman (1962) and Lorenz (1963), we perform a truncated Fourier expansion of $\hat{\psi}$ and $\hat{\theta}$, assuming that they are periodic along the $\hat{x}$ and $\hat{z}$ with periodicity of $2/a$ and $2$, respectively. When boundary conditions are considered, we then derive a set of ODEs describing the temporal evolution of the corresponding (complex valued) modes $\Psi_{m,n}$ and $\Theta_{m,n}$, which are associated to the wavevector $\vec{K} = (\pi a n, \pi m)$. See Saltzman (1962) for a detailed derivation in the case $Ec = 0$, considering that the sign of the term proportional to $R$ is wrong in both Eqs. 34 and 35. In the case $Ec = 0$, the seminal Lorenz system can be derived by severely truncating the system, considering the evolution equation for the real part of $\Psi_{1,1}$ and for the imaginary parts of $\Theta_{1,1}$ and $\Theta_{0,2}$, and performing suitable rescaling (see below). While $\Theta_{0,2}$ has no real part because of the boundary conditions (Saltzman, 1962), neglecting the imaginary (real) part of $\Psi_{1,1}$ ($\Theta_{1,1}$) amounts to an arbitrary selection of the phase of the waves in the system. An entire hierarchy of generalized Lorenz models, all obeying to this constraint, can be derived with lengthy but straightforward calculations. See, *e.g.*, Curry (1978), Wenyi and Peicai (1984), Roy and Musielak (2007) for detailed discussion of these models.

**Symmetries of the extended Lorenz system**

In this work we include the modes $\Psi_{1,1}$, $\Psi_{2,2}$, $\Theta_{1,1}$, $\Theta_{2,2}$ in our truncation, and retain both the real and the imaginary parts, whereas the considered horizontally symmetric modes $\Theta_{0,2}$ and $\Theta_{0,4}$ are, as mentioned above, imaginary. Finally, we assume, in general, a non vanishing value for $Ec$. If we define $\Psi_{1,1} = \alpha(X_1 + iX_2)$, $\Theta_{1,1} = \beta(Y_1 + iY_2)$, $\Psi_{2,2} = \alpha(A_1 + iA_2)$, $\Theta_{2,2} = \beta(B_1 + iB_2)$, $\Theta_{0,2} = i\gamma Z_1$, and $\Theta_{0,4} = i\gamma Z_2$, with $\alpha = (a^2+1)/(2\sqrt{2}a)$, $\beta = \pi^3(a^2+1)^3/(2\sqrt{2}a^2)$, $\gamma = \pi^3(a^2+1)^3/(2a^2)$, set $a = 1/\sqrt{2}$ (Lorenz, 1963), we derive the following system of real ODEs:

$$\dot{X}_1 = \sigma Y_2 - \sigma X_1$$
$$\dot{X}_2 = -\sigma Y_1 - \sigma X_2$$
$$\dot{Y}_1 = X_2 Z_1 - rX_2 - Y_1 + 5/(\sqrt{2}\pi)\sigma Ec(A_1 X_1 + A_2 X_2)$$
$$\dot{Y}_2 = -X_1 Z_1 + rX_1 - Y_2 + 5/(\sqrt{2}\pi)\sigma Ec(A_2 X_1 - A_1 X_2)$$
$$\dot{A}_1 = \sigma/2\, B_2 - 4\sigma A_1$$
$$\dot{A}_2 = -\sigma/2\, B_1 - 4\sigma A_2 \quad\quad (3a\text{-}3l)$$
$$\dot{B}_1 = 4A_2 Z_2 - 2rA_2 - 4B_1 - 1/(2\sqrt{2}\pi)\sigma Ec(X_1^2 - X_2^2)$$
$$\dot{B}_2 = -4A_1 Z_2 + 2rA_1 - 4B_2 - 1/(\sqrt{2}\pi)\sigma Ec X_1 X_2$$
$$\dot{Z}_1 = X_1 Y_2 - X_2 Y_1 - bZ_1$$
$$\dot{Z}_2 = 4A_1 B_2 - 4A_2 B_1 - 4bZ_2$$

where the dot indicates the derivative with respect to $\tau = \pi^2(a^2+1)t = 3/2\pi^2 t$, $r = R/R_c = R/(27\pi^4/4)$ is the relative Rayleigh number, and $b = 8/3$ is a geometric factor. Note that if we exclude the faster varying modes $\Psi_{2,2}$ and $\Theta_{2,2}$ in our truncation (as in the Lorenz case) the Eckert number is immaterial in the equations of motion, the basic reason being that viscous dissipation acts on small spatial scales. Therefore, the ODEs (3a-3l) provide the minimal system which includes the feedback due to the thermal effect of viscosity.

If we retain all the variables in the system (3a-3l) and $Ec$ is set to 0, the following symmetry is obeyed. Let $S(t, S_0) = (X(t), Y(t), A(t), B(t), Z(t))^T$ - with $\bullet(t) = (\bullet_1(t), \bullet_2(t))$ - be a solution of the system with initial conditions $S_0 = (X(0), Y(0), A(0), B(0), Z(0))^T$. We then have that:

$$T^{-1}(\omega,\phi)S(t,T(\omega,\phi)S_0) = S(t,S_0) \qquad (4)$$

for all $(\omega,\phi) \in \Re^2$, where $T(\omega,\phi)$ is a real, linear, unitary ($T^{-1}(\omega,\phi) = T^T(\omega,\phi)$) transformation:

$$T(\omega,\phi)S = \begin{pmatrix} R_\omega & 0 & 0 & 0 & 0 \\ 0 & R_\omega & 0 & 0 & 0 \\ 0 & 0 & R_\phi & 0 & 0 \\ 0 & 0 & 0 & R_\phi & 0 \\ 0 & 0 & 0 & 0 & I \end{pmatrix} \begin{pmatrix} X \\ Y \\ A \\ B \\ Z \end{pmatrix}, \qquad (5)$$

with $I$ being the 2X2 identity matrix and $R_\xi$ being the 2X2 rotation matrix of angle $\xi$. The $T(\omega,\phi)$-matrix shifts the phases of both the $\Psi_{1,1}$ and $\Theta_{1,1}$ waves by of the same angle $\omega$ and, independently, the phases of both the $\Psi_{2,2}$ and $\Theta_{2,2}$ waves by the same angle $\phi$. The symmetry (4-5) gives a more solid framework to the result by Chen and Price (2006), generalizes their approach, and shows that the dynamics of the variables $(X_1, X_2, Y_1, Y_2, Z_1)$ is entirely decoupled from that of the variables $(A_1, A_2, B_1, B_2, Z_2)$

**Statistical properties of the system – symmetric case**

In the $Ec = 0$ case, we expect the presence of degeneracies in the dynamics, leading to non-ergodicity of the system. Since, from (4), we have that $S(t,T(\omega,\phi)S_0) = T(\omega,\phi)S(t,S_0)$, we readily obtain that the statistical properties of the flow depend on the initial conditions, and that by suitably choosing the matrix $T(\omega,\phi)$, we can, e.g., exchange the statistical properties of $X_1$ and $Y_1$ with those of $X_2$ and $Y_2$ by setting $\omega = \pi/2$ (and similarly for the $A$ and $B$ variables by setting $\phi = \pi/2$). Whereas, in actual integration, numerical noise coupled with sensitive dependence on initial conditions (see below) breaks the instantaneous identity $S(t,T(\omega,\phi)S_0) = T(\omega,\phi)S(t,S_0)$ after sufficiently long time, the statistical properties of the observables transform according to the symmetry defined in Eq. (5).

With the "classical" Lorenz parameter values $r = 28$, $\sigma = 10$, $b = 8/3$, the variables $(A_1, A_2, B_1, B_2, Z_2)$ describe the faster spatially varying wave components; they do not feature any time variability when asymptotic dynamics is considered, as they converge - rather quickly – to fixed values $(A_1^\infty, A_2^\infty, B_1^\infty, B_2^\infty, Z_2^\infty)$. Since the symmetry defined in Eq. (5) is obeyed, $\left(A_1^\infty, A_2^\infty, B_1^\infty, B_2^\infty\right)$ depend on the initial conditions, while the quadratic quantities $\left|A^\infty\right|^2 = \left(A_1^\infty\right)^2 + \left(A_2^\infty\right)^2$, $\left|B^\infty\right|^2 = \left(B_1^\infty\right)^2 + \left(B_2^\infty\right)^2$, and $Z_2^\infty$, do not. Additionally, the *curl-like* quantity $A_1^\infty B_2^\infty - A_2^\infty B_1^\infty$, linked to the relative rotation of the *A* and *B* variables, does not depend on the initial conditions. After some algebraic manipulations, we obtain $\left|A^\infty\right|^2 = b/8(r/2 - 8)$, $\left|B^\infty\right|^2 = 8b(r/2 - 8)$, $Z_2^\infty = r/2 - 8$, and $A_1^\infty B_2^\infty - A_2^\infty B_1^\infty = b(r/2 - 8)$.

Instead, the variables $\left(X_1, X_2, Y_1, Y_2, Z_1\right)$ have an erratic evolution and their dynamics defines an extended Lorenz system (Chen and Price, 2006). The classical three-component system results from a specific selection of the phase of the waves of the system, which is obtained by setting vanishing initial conditions for $X_2$ and $Y_1$. As a consequence of the symmetry given in Eq. (5), the statistical properties of $Z_1$ do not depend on the initial conditions, and agree with those of the *Z* variable of the classical Lorenz system. Moreover, the statistical properties of the quadratic quantities $\left|X\right|^2 \equiv X_1^2 + X_2^2$, $\left|Y\right|^2 \equiv Y_1^2 + Y_2^2$, and $\left|XY\right| \equiv X_1 Y_2 - X_2 Y_1$ do not depend on the initial conditions (whereas those of each term is the previous sums do!), and agree with those of $X^2$ and $Y^2$, and *XY* of the classical Lorenz system, respectively.

From a physical point of view, the symmetry (4-5) is related to the fact that the system does not preferentially select streamfunction and temperature waves of a specific phase, and, does not mix phases. Therefore, in a statistical sense, the relative strength of waves with the same periodicity but different phase can be arbitrarily chosen by suitably selecting the initial conditions. Instead, the statistical properties of the space-averaged convective heat transport, which is determined by a linear combination of $Z_1$ and $Z_2$, as well as those of the total kinetic energy (determined by a linear combination of $\left|X\right|^2$ and $\left|A\right|^2$) and available potential energy (determined by a linear combination of $\left|Y\right|^2$ and $\left|B\right|^2$) of the fluid (Saltzman, 1962) must not depend on the initial conditions - but only on the system's parameters - as they are robust (thermo-)dynamical properties of the turbulent flow.

The attractor of the system (3a-3l) can be expressed as a Cartesian product of a 2-thorus - as given by the rotational symmetry (4-5) - with a "fundamental" strange attractor. Therefore, despite the fact that ergodicity is not obeyed, it makes sense to compute the Lyapunov exponents (Osedelec, 1968; Ruelle, 1989) of the system (3a-3l). Actually, if we compute the Lyapunov exponents $\left(\lambda_1, ..., \lambda_{10}\right)$ using the algorithm by Benettin et al. (1980), we obtain the same results independently of the initial conditions. The sum of the Lyapunov exponents is given by the trace of the jacobian *J* of the system (3a-3l) and yelds $Tr(J) = \sum \lambda_j = -5(2 + 2\sigma + b) \approx -123.333$. In particular, we obtain one positive exponent ($\lambda_1 \approx 0.905$), which gives the metric entropy $h = \sum_{\lambda_j > 0} \lambda_j$ of the system (Ruelle, 1989) and matches exactly the same value as the positive Lyapunov exponent in the classical Lorenz system, and three vanishing exponents $\lambda_2 = \lambda_3 = \lambda_4 = 0$. The non-hyperbolicity of the system, as described by the presence of two additional vanishing exponents, is related to the existence of two neutral directions due to the symmetry property (4-5). Moreover, we have 6 negative Lyapunov exponents, one of which agrees with the negative exponent of the Lorenz system ($\lambda_8 \approx -14.572$). Given the very structure (and derivation) of the model discussed above, the fact that we recover the classical results of the 3-component Lorenz system is quite reinsuring. The Kaplan-Yorke dimension (Kaplan and Yorke, 1979) of the system is:

$$d_{KY} = k + \frac{\sum_{j=1}^{k} \lambda_j}{|\lambda_{k+1}|} \approx 4.183 \qquad (6)$$

where *k* is such that the sum of the first *k* (4, in our case) Lyapunov exponents is positive and the sum of the first *k*+1 Lyapunov exponents is negative.

**Symmetry-break and phase mixing**

Considering a non-vanishing value for $Ec$ amounts to including an additional coupling between the temperature and the streamfunction waves. Such a coupling breaks the symmetry (4-5), so that a dramatic impact on the attractor properties is observed.

When very large values of $Ec$ are considered ($Ec > 0.045$), the system loses its chaotic nature as no positive Lyapunov exponents are detected (not shown), whereas periodic motion is realized. The analysis of this transition and of this regime is beyond the scopes of this paper, as we confine ourselves to studying the properties of the system when chaotic motion is realized, thus focusing on the $Ec \to 0$ limit, thus considering $Ec \in [0, 0.02]$.

In physical terms, the coupling allows for a mixing of the phases of the thermal and streamfunction waves, thus destroying the previously described degeneracies and establishing ergodicity in the system. Note that, since also in the case of $Ec > 0$ the convection does not preferentially act on waves of a specific phase, we obtain that, pairwise, the statistical properties of $X_1$ and $X_2$, and of $Y_1$ and $-Y_2$ are identical and do not depend on the initial conditions. The same applies for the $A$ and $B$ pairs of variables, respectively. If, in particular, we consider the long term averages of the $\langle |X|^2 \rangle$, $\langle |Y|^2 \rangle$, $\langle |XY| \rangle$, $\langle |A|^2 \rangle$, $\langle |B|^2 \rangle$, and $\langle |AB| \rangle$, we obtain that in all cases the two addends give the same contribution, e.g. $\langle X_1^2 \rangle = \langle X_2^2 \rangle = 1/2 \langle |X|^2 \rangle$. In this case, long is considered with respect to the additional time scale $Ec^{-1}$ introduced when taking into account the thermal effect which viscosity introduces into the system. Given an initial condition, after a time of the order of $Ec^{-1}$ time units the system "realizes" that the symmetry (4-5) is broken and mixing of the phases of the waves becomes substantial. Note that, as $Ec^{-1}$ is much larger than the other time scales, we deal with a stiff system.

Whereas having $Ec > 0$ is crucial in terms of mixing the phases of waves, the impact of the symmetry break on the long term averages of the physically sensitive observables $O = |X|^2$, $|Y|^2$, $|A|^2$, $|B|^2$, $Z_1$, and $Z_2$ is relatively weak in the range of values $Ec \in [0, 0.02]$, as in all cases we have that $\Delta \log(\langle O \rangle) \approx Ec$. In particular, we obtain that with increasing values of $Ec$ thermal waves ($\langle |Y|^2 \rangle, \langle |B|^2 \rangle$) are enhanced, faster spatially varying streamfunction waves ($\langle |B|^2 \rangle$) become stronger at the expense of waves with slower spatially varying waves ($\langle |X|^2 \rangle$), and finally, the contribution to the convective heat transport due to the faster spatially varying mode ($\langle Z_2 \rangle$) increases at the expense of the slower mode ($\langle Z_1 \rangle$).

**Symmetry- break and hyperbolicity**

We hereby want to take a different point of view on the process of symmetry break of the dynamical system and on its sensitivity with respect to $Ec$, by analyzing the $Ec$-dependence of the spectrum of the Lyapunov exponents. The first, crucial result system is that the system is hyperbolic for any finite positive value of $Ec$: two of the vanishing Lyapunov exponents branch off from 0 with a distinct linear dependence on $Ec$, so that $\lambda_2 \approx 5.1 Ec$ and $\lambda_4 \approx -6.0 Ec$ for $Ec < 0.008$, whereas $\lambda_3 = 0$ corresponds to the direction of the flow. The largest Lyapunov exponent also decreases linearly as $\lambda_1 \approx 0.905 - 9.1 Ec$, whereas all the other Lyapunov exponents feature a negligible dependence on $Ec$, except $\lambda_8 \approx -14.572 + 10.0 Ec$, which ensures that the sum of Lyapunov exponents is independent of $Ec$. Therefore, we derive that $Ec < 0.008$ the metric entropy decreases with $Ec$ as $h(Ec) = \lambda_1 + \lambda_2 \approx 0.905 - 4.0 Ec$, whereas the Kaplan-Yorke dimension can be parameterized as $d_{KY}(Ec) = 4 + (\lambda_1 + \lambda_2 + \lambda_3 + \lambda_4)/|\lambda_5| \approx 4.183 - 2.01 Ec$. For larger values of $Ec$, linearity is not obeyed (except regarding the $Ec$-dependence $\lambda_2$ and $\lambda_4$),

whereas a faster, monotonic decrease of both the metric entropy and the Kaplan-Yorke dimensions are found. See Figs. 1a)-1c) for results for values of $Ec$ up to 0.02. We then conclude that, in spite of introducing a second unstable direction, which is responsible for mixing the phases of the waves, the inclusion of the impact of the viscous dissipation on the thermal energy balance acts with continuity on the dynamical indicators, by reducing the overall instability and increasing the predictability of the system, as well as confining the asymptotic dynamics to a more limited (in terms of dimensionality) set.

**Symmetry-break and low-frequency variability**

We now focus on higher-order statistical properties of the system. We first observe that the variables $A_1$, $A_2$, $B_1$, $B_2$, $Z_2$ feature mostly ultra-low frequency variability. In order to provide some qualitative highlights on these variables, in Fig. 2 we show, from top to bottom, the projection on $A_1$ of three typical trajectories for and $Ec = 10^{-3}$, $Ec = 10^{-4}$, and $Ec = 10^{-5}$, respectively. Going from the top via the middle to the bottom panel, the time scale increases by a factor of 10 and 100, respectively. The striking geometric similarity underlines that the time scales of the dominating variability for the slow variables can be estimated as $1/\lambda_2 \approx Ec^{-1}$. As we take the $Ec \to 0$ limit, such a time scale goes to infinity, which agrees with the fact that for $Ec = 0$ these variables converge asymptotically to fixed values. Moreover, as $A_1^2 + A_2^2 \approx b/8(r/2 - 8) = 2$ when $Ec^{-1}$-long time averages are considered, Fig. 1 implies that the system switches (on short time scales) back and forth between $A_1$- and $A_2$-dominated dynamics. The same observation are made for the $B_1$ and $B_2$ pair of variables.

On the other hand, the dynamics of $X_1$, $X_2$, $Y_1$, $Y_2$, $Z_1$ is basically controlled by the time scale $1/\lambda_1 \ll 1/\lambda_2$, so that a clear-cut separation between fast and slow variables can be figured out and quantitatively justified. In Fig. 3 we depict, from the top to bottom panel, some trajectories of the fast variable $X_1$ for $Ec = 10^{-3}$, $Ec = 10^{-4}$, and $Ec = 10^{-5}$, respectively. Similarly to Fig. 2, the time scale of the x-axis is scaled according to $Ec^{-1}$. Since $X_1$ is a fast variable, the dominating high-frequency variability component is only barely affected by changing $Ec$, because $\lambda_1$ has a weak dependence on $Ec$, as discussed in the previous section (see also Fig. 1a). Nevertheless, in Fig. 3 we observe an amplitude modulation occurring on a much slower time scale of the order of $1/\lambda_2 \approx Ec^{-1}$. Note, as in Fig. 2, that such a slow dynamics of the "coarse grained" $X_1$ variable for various values of $Ec$ is statistically similar when the time is scaled according to $Ec^{-1}$. As in the previous case, the slow amplitude modulation determines the changeovers between extended periods of $X_1$– and $X_2$–dominated dynamics, thus ensuring the phase mixing and ergodicity of the system. In the $Ec \to 0$ limit, as discussed above, the relative strength of the two phase components of the (1,1) waves is determined by the initial conditions, as mixing requires $\approx Ec^{-1}$ time units. The same considerations apply for the $Y_1$ and $Y_2$ pair of variables.

Further insight can be obtained by looking at the spectral properties of the fast and slow variables; some relevant examples of power spectra for $Ec = 10^{-3}$, $Ec = 10^{-4}$, and $Ec = 10^{-5}$ are depicted in Fig. 4, where we have considered the low frequency range by concentrating on time scales larger than $10/\lambda_1$. The white noise nature of the fast variable $X_1$ is apparent and so is the overall lack of sensitivity of its spectral properties with respect to $Ec$. Note that any signature of the amplitude modulation, which is responsible for phase mixing observed in Fig. 3, is virtually absent. This shows how crucial physical processes are masked, due to their weakness, when using a specific *metric*, like that provided by the power spectrum. Instead, the slow variable $A_1$ has a distinct red power spectrum, which features a dominating $f^{-3/2}$ scaling in the low frequency range (black line) between the time scales $\approx 0.5\ Ec^{-1}$ and $\approx 10\ Ec^{-1}$. Such a scaling regime dominates the ultra-low frequency variability described in Fig. 2 and is responsible for the long-term memory of the signal. For higher frequencies, the spectral density $\rho(P)$ decreases - see the sharp corner in the for $f \approx 0.5Ec$ - as $f^{-4}$, and then a classical red noise spectrum $\propto f^{-2}$ is realized for very low values of the spectral density. Note also that, in agreement with our visual perception of Fig. 2, the spectral density $\rho(P)$ scales according to $Ec^{-1}$ in a large range of energy-containing time scales.

**Summary and Conclusions**

In this work we have derived and thoroughly analyzed a 10-variable system of Lorenz-like ODEs obtained by a severe spectral truncation of the 2D x-z convective equations for the streamfunction $\psi$ and temperature $\theta$ in Boussinesq approximation and subsequent nondimensionalization. in the truncation we retain the terms corresponding to the real and imaginary part of the modes $\Psi_{m,n}$ (streamfunction) and $\Theta_{m,n}$ (temperature) associated to the *x-z* wavevector $\vec{K} = (\pi an, \pi m)$ with $(m,n) = (1,1)$ and $(m,n) = (2,2)$, and the imaginary part of the x-symmetric modes $\Theta_{0,2}$ and $\Theta_{0,4}$. As modes characterized by a faster varying spatial structure and different parity are added to the Lorenz spectral truncation, which describes the dynamics of the real part of $\Psi_{1,1}$ and of the imaginary part of $\Theta_{1,1}$ and $\Theta_{0,2}$, the system represents the thermal impact of viscous dissipation, controlled by the Eckert number *Ec*. The presence of a forth parameter - together with the usual

Lorenz parameters $r$ (relative Rayleigh number), $\sigma$ (Prandl number), and $b$ (geometric factor) – marks a crucial difference between this ODEs system and other extensions of the Lorenz system proposed in the literature - see, *e.g.*, Curry (1978), Wenyi and Peicai (1984), Roy and Musielak (2007). Moreover, as can be deduced following the argumentations of Nicolis (1999), this system specifically allows for the closed-form computation of the entropy production, as the Eckert number enters into its evaluation.

The following results are worth mentioning:

1. When the Eckert number is set to 0, as common in most applications, the dynamical system is invariant with respect to the action of a specific symmetry group, which basically shifts the phases of the streamfunction and thermal waves with $m > 0$. At a physical level this implies that the initial conditions set the relative strength of waves with the same spatial structure but π/2 –shifted phases. These results extend the findings by Chen and Price (2006), which considered a 5-variable system. In particular, it is found that the Lorenz system is included in the ODEs system analyzed here when specific initial conditions, which set the parity of the slower spatially varying modes, are selected. The absence of phase-mixing implies a lack of ergodicity, and the degeneracy of the dynamics is reflected also in the non-hyperbolicity of the system. When the classical parameters' values $r = 28$, $\sigma = 10$, $b = 8/3$ are selected, three of the Lyapunov exponents vanish, one corresponding to the direction of the flow, the other two accounting for the toroidal symmetry. The value of the only positive exponent coincides with that of the Lorenz system, and the value of one of the six negative exponents agrees with that of the negative Lyapunov exponent of the Lorenz system. Therefore, the Lorenz system contains already all the interesting unstable dynamics described by this extended ODEs system, and features exactly the same value for the metric entropy. Correspondingly, while the five variables (*fast*) describing the modes $\Psi_{1,1}$, $\Theta_{1,1}$, and $\Theta_{0,2}$ have an erratic behavior, the other five variables (*slow*) converge to fixed values.

2. When $Ec \neq 0$, the symmetry of the system is broken, and coupling occurs between the fast and slow variable. If we select,a sin the original Lorenz system, $r = 28$, $\sigma = 10$, $b = 8/3$ and, the system is chaotic for $0 < Ec \leq 0.045$, whereas for higher $Ec$-values a quasi-periodic regime is realized. In the chaotic regime, the symmetry-break is accompanied by the establishment of a hyperbolic dynamics: two Lyapunov exponents branch off from zero (one positive, one negative) linearly with $Ec$. Therefore, a second unstable direction with a second time scale $1/\lambda_2 >> 1/\lambda_1$ with $1/\lambda_2 \approx Ec^{-1}$ is established. Overall, the impact of the thermal-viscous feedback is stabilizing, as indicated by the metric entropy and the Kaplan-Yorke attractor dimension monotonically decreasing with increasing $Ec$, with a marked linear behavior for $Ec \leq 0.008$.

3. The coupling establishes dynamics on time scales of the order of $Ec^{-1}$ responsible for the changeovers between extended periods of dominance of waves of specific phase, both for the slow and for the fast variables. Such dynamical processes, which result from small terms in the evolution equations, correspond to the mixing of phases of the waves and ensures the ergodicity of the system. In particular, the slow variables have a non-trivial time-evolution and are characterized by a dominating $f^{-3/2}$ scaling in the low frequency range for time scales between $0.5\ Ec^{-1}$ and $10\ Ec^{-1}$. Instead, in the case of fast variables, the phase mixing appears as a slow amplitude modulation occurring on time scales of $Ec^{-1}$ which superimposes on the fast dynamics controlled by the time scale $1/\lambda_1$.

The system introduced in this paper features very rich dynamics and, therefore, may have prototypical value for phenomena generic to complex systems, such as the interaction between slow and fast variables and the presence of long term memory. Moreover, analysis shows how, neglecting the coupling of slow and fast variables only on the basis of scale analysis – as usually done when discarding the Eckert number - can be catastrophic. In fact, this leads to spurious invariances that affect essential dynamical properties (ergodicity, hyperbolicity) and that cause the model losing its ability to describe intrinsically multiscale processes. This may suggest that a careful re-examination of the scaling procedures commonly adopted for defining simplified models, especially in the climate science community, may be fruitful in the development of more efficient modeling strategies.

We can point at four possible future lines of research:

1. extension of the $Ec = 0$ invariance properties of similarly obtained, higher order truncation ODEs systems, and the impact on these symmetries resulting from setting $Ec > 0$;

2. analysis of the long-term memory of the slow variables and of the related statistical properties of extreme events in terms of the $Ec$-dependence of GEV parameters (Coles, 2001; Felici et al., 2007);

3. investigation of how predictability properties of the system depend on $Ec$: is there a range of values of $Ec$ without exhibiting the unrealistic return of forecast accuracy as shown in several low dimensional systems, see *e.g.* Smith et al. (1999)?

4. evaluation of the entropy production of the system, along the lines envisioned by Nicolis (1999).

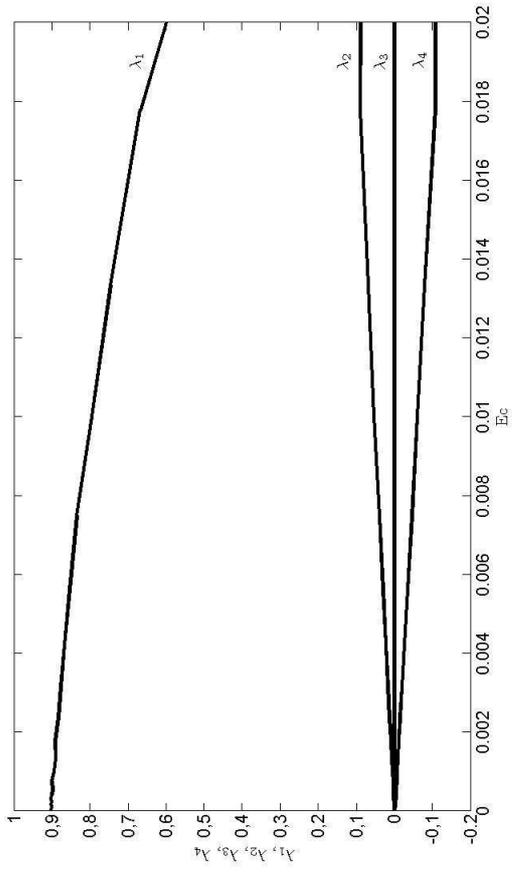 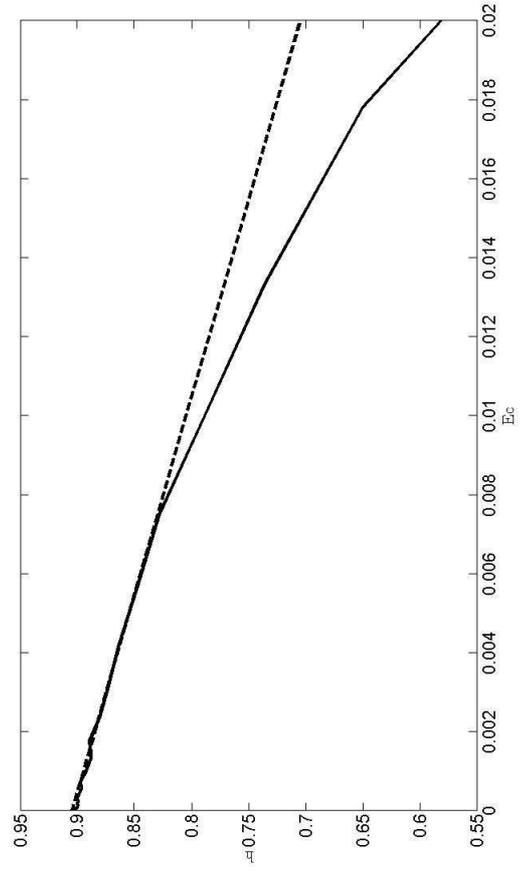

(a) (b)

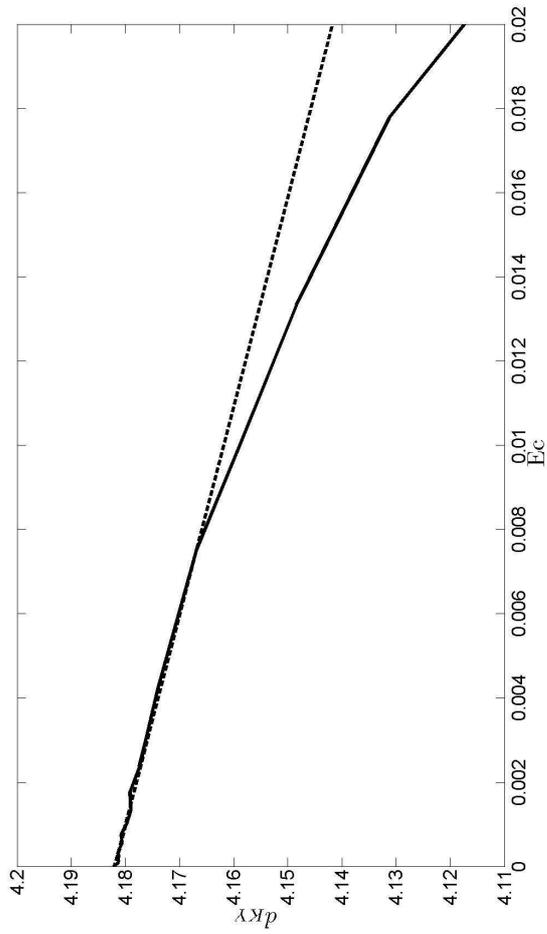

(c)

**Figure 1: Four largest Lyapunov exponents (a), metric entropy $h$ (b) and Kaplan-Yorke $d_{KY}$ dimension (c) as a function of $Ec$. Note the continuity for $Ec = 0$ of all parameters and the distinct linear behavior - see specifically the dashed lines in (b) and (c) - for $Ec < 0.008$. The linear behavior of the second and fourth Lyapunov exponents branching off zero in (a) extends throughout $Ec \approx 0.018$. Details in the text.**

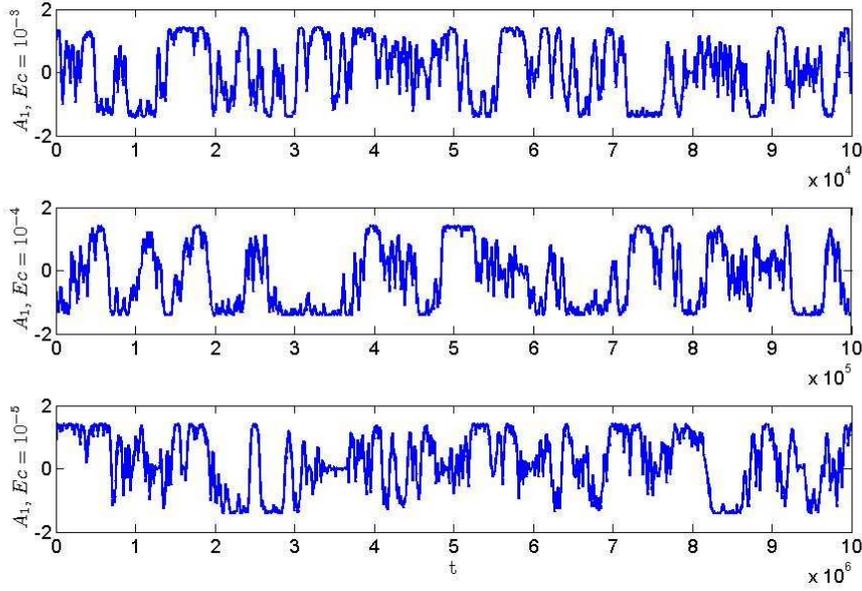

**Figure 2: Impact of the viscous-thermal feedback on the time scales of the system. From top to bottom: typical evolution of the variable $A_1$ for different values of the Eckert number ($Ec = 10^{-3}$, $Ec = 10^{-4}$, $Ec = 10^{-5}$, respectively). Note that the time scale is magnified by a factor of 1, 10 and 100 from top to bottom. Details in the text.**

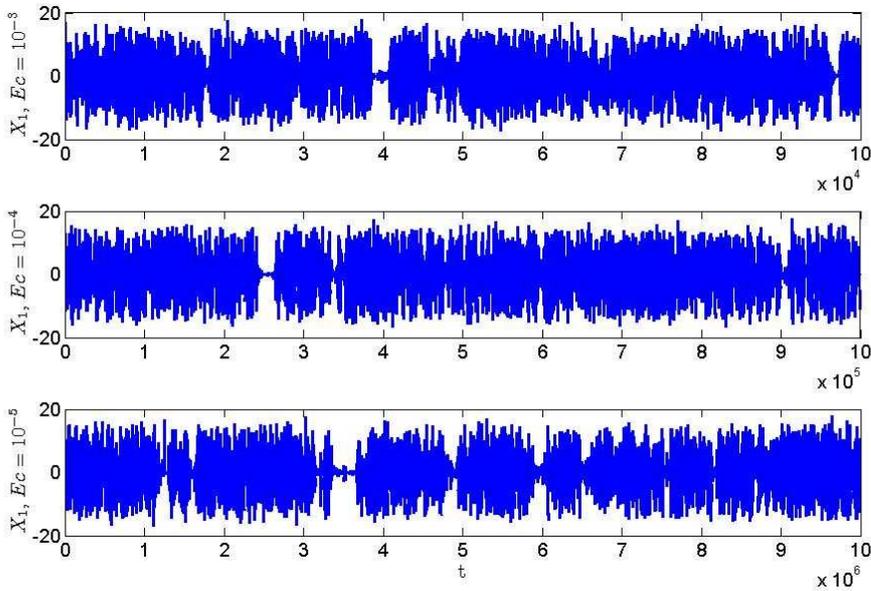

**Figure 3: Impact of the viscous-thermal feedback on the time scales of the system. From top to bottom: typical evolution of the variable $X_1$ for different values of the Eckert number ($Ec = 10^{-3}$, $Ec = 10^{-4}$, $Ec = 10^{-5}$, respectively). Note that the time scale is magnified by a factor of 1, 10 and 100 from top to bottom. Details in the text.**

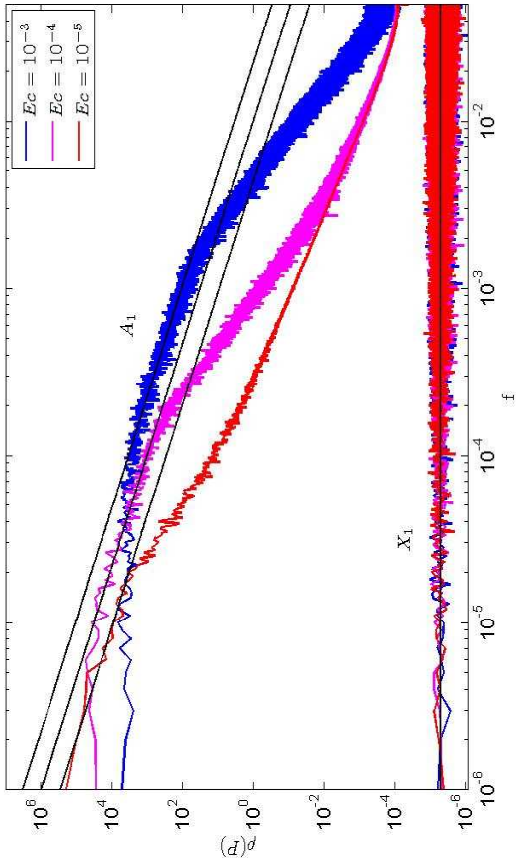

**Figure 4: Power spectrum of $A_1$ and $X_1$ (multiplied times $10^{-8}$) for $Ec = 10^{-3}$, $Ec = 10^{-4}$, and $Ec = 10^{-5}$ in units of power per unit frequency. The black lines correspond to $f^{-3/2}$ scalings. Details in the text.**

# Bibliography

E. N. Lorenz, J. Atmos. Sci. 20, 130 (1963)

D. Ruelle, *Chaotic Evolution and Strange Attractors*, Cambridge University Press, Cambridge, 1989

B. Saltzman, J. Atmos. Sci. 19, 329 (1962)

J. H. Curry, Commun. Math. Phys. **60**, 193 (1978).

Z. Wenyi and Y. Peicai, Adv. Atmos. Sci. **3**, 289 (1984)

D. Roy and Z.E. Musielak, Chaos, Solitons and Fractals **31** (2007) 747

C. Sparrow, *The Lorenz Equations. Bifurcations, Chaos, and Strange Attractors*, Springer, New York, 1984

C. Bonatti, L. J. Diaz, and M. Viana, *Dynamics Beyond Uniform Hyperbolicity: A Global Geometric and Probabilistic Perspective*, Springer, New York, 2005

V. Lucarini, J. Stat. Phys. **131**, 543 (2008a)

D. Ruelle, Nonlinearity, **11** (1998), 5. 235

V. Lucarini, J. Stat. Phys., in press (see also: arxiv:condmat/0809.0101v1) (2008b)

L.A. Smith, C. Ziehmann, and K. Fraedrich, Q. J. R. Meterol. Soc. **125**, 2855 (1999)

T.N. Palmer, Bull. Am. Met. Soc. **74**, 49 (1993)

C. Nicolis, Quart. J. Roy. Meteor. Soc. **125,** 1859 (1999)

Z.-M. Chen and W.G. Price, Chaos, Solitons and Fractals **28** 571 (2006)

V. Oseledec, Moscow Math. Soc.**19**, 197 (1968)

G. Benettin, L. Galgani, A. Giorgilli and J. M. Strelcyn, Meccanica **15**, 9 (1980)J. L. Kaplan and J. A. Yorke, in H.O. Peitgen and H.O. Walther: *Functional Differential Equations and Approximations of Fixed Points*. Springer, New York, 1979

S. Coles, *An Introduction to Statistical Modeling of Estreme Values*. Springer, Heidelbeg, 2001

M. Felici, R. Vitolo, V. Lucarini, and A. Speranza, J. Atmos. Sci. **64**, 2137 (2007)